\documentclass[12pt,a4paper,final]{iopart}
\usepackage[latin1]{inputenc}
\usepackage[T1]{fontenc}
\usepackage{bm}
\usepackage{graphicx}
\expandafter\let\csname equation*\endcsname\relax
\expandafter\let\csname endequation*\endcsname\relax
\usepackage{amsmath}
\usepackage{amsfonts}
\usepackage{amssymb}
\usepackage{color}
\usepackage{verbatim}
\usepackage{citesort}
\begin{document}
\title[Charge transport]{Charge transport through a semiconductor quantum dot-ring nanostructure}
\author{Marcin~Kurpas$^1$, Barbara~K\k{e}dzierska$^1$, Iwona~Janus-Zygmunt$^1$, Anna~Gorczyca-Goraj$^1$, 
El\.zbieta~Wach$^2$, El\.zbieta Zipper$^1$ and Maciej~M~Ma\'{s}ka$^1$}
\address{$^1$Department of Theoretical Physics, University of Silesia, Uniwersytecka 4, 40-007 Katowice, Poland }
\address{$^2$Faculty of Physics and Applied Computer Science, AGH University of Science and Technology, 
Mickiewicza 30, 30-059 Krak\'ow, Poland }
\ead{maciej.maska@us.edu.pl}

\begin{abstract}
Transport properties of a gated nanostructure depend crucially on the coupling of its states to the states of electrodes. 
In the case of a single quantum dot the coupling, for a given quantum state, is constant or can be slightly modified by additional
gating. In this paper we consider a concentric dot--ring nanostructure (DRN) and show that its transport properties can be 
drastically modified due to the unique geometry. We calculate the dc current through a DRN in the Coulomb blockade regime and show 
that it can efficiently work as a single electron transistor or a current rectifier. In both cases the transport characteristics 
strongly depends on the details of the confinement potential. The calculations are carried out for low and high bias regime, the latter 
being especially interesting in the context of current rectification due to fast relaxation processes. 
\end{abstract}
\maketitle 

%
\section{Introduction}\label{sec1}
In order to meet growing demand for small scale, low-power consuming devices one has to downscale transistors and 
logic circuits and to work with a small number of carriers. The natural limit for lowering carrier density is single charge electronics,
where phenomena such as electric current can be controlled with single electron precision \cite{kouwen_1991,vdVaart_1995}.
Contrary to modern mass production electronics, single and a few electron devices exhibit purely quantum mechanical effects such as 
resonant tunneling \cite{kouwen_1991,devoret_1992,kastner_1992,vdVaart_1995,hans} or quantum entanglement \cite{blais_2004,petta_2005,versteegh_2014}. Apart from direct applications in
nanoelectronics they are also perfect tools for probing fundamental problems in single and many-body physics.
During the last decade a lot of research has been devoted to study the electronic properties of qauntum dots (QD) 
\cite{hans,enslin,Fudzi,amasha,fuhrer,scheib,lis}. 
For sufficiently low temperatures the discretness of the energy spectrum  of these systems can be clearly 
visible in transport experiments as Coulomb peaks \cite{averin_1986,kouwen_1991,kastner_1992,coulomb_blockade} that demonstrate succesive charging and discharging 
of the QD by single electrons. The Coulomb blockade phenomenon, where the charging energy forbids an electron to jump to the 
QD, is the basis for most of the applications  of QDs \cite{fuji_2002,koppens_2006,elzerman_2004}.
Quantum dots arranged into double, triple or more complex systems 
\cite{weber,shinkai,McNeil,hatano,ota,koppens,stopa,StopaVidan,hawrylak2} exhibit abundance of quantum states which manifest 
themselves, e.g., in Pauli spin blockade \cite{Ono,spin_blockade_2} current or heat rectification \cite{scheib} effects. 

Another class of interesting quantum systems are quantum rings (QR) \cite{zkm,kurpas_2014,piacente,lei,kleemans}. Due to different from 
QDs geometry the phenomena observed in QRs are very sensitive to phase coherence of the electronic wave function. These are, e.g., the 
Aharonov-Bohm effect demonstrating modification of the electron wave function by a vector potential \cite{A_B} or persistent currents, i.e., 
ground state currents that flow in QR even without an external magnetic field \cite{PC_1,PC_2,PC_3}.

In this paper we focus on a complex nanostructure that combines the two mentioned above, topologically different components: 
a quantum dot and a quantum ring.  The constituents are aligned concentrically (QD is surrounded by QR) so that the system 
conserves the circular symmetry. The dot-ring nanostructure (DRN) has already been fabricated by pulsed droplet epitaxy 
\cite{somaschini,sangu} with full control of the growth process. It can also be made by using atomic force microscope to 
locally oxidize the surface 
of a sample \cite{fuhrer} or by lithography. Another method would be to grow a core-shell nanowire 
 \cite{lauhon,dillen} 
of, e.g., (In,Ga)As, where the core and shell parts are separated by a tunneling barrier. Then by cutting a slice of it one 
can form a DRN.  A DRN exhibits a large variety of quantum states \cite{peeters} what leads to many interesting features. 

It has been recently shown \cite{zipper,KurpasKedz} that many measurable properties of a DRN, like spin relaxation or optical
absorption, can be widely changed by a modification of the confinement potential of the DRN demonstrating its very high 
controllability and flexibility. These characteristics are mostly determined by the relative distribution of the wave functions
in a DRN that, in turn, can be changed by external gates or fields. 
The purpose of this study is to demonstrate that also conducting properties of a DRN are very sensitive to the details of 
confinement and that we can realize different single electron devices on this complex structure. 
Unlike field effect transistors, single electron devices are based on intrinsically quantum phenomenon, namely the tunnel
effect. In this case transport properties are mostly determined by the tunneling rates $\Gamma$'s, which depend on the overlap
of the DRN states with the states of the electrodes.
These parameters, in turn, depend crucially on the localization of the electron wave
function: states localized in QD (QR) are weakly (strongly) coupled to the electrodes. Thus $\Gamma$'s may strongly
 depend on the quantum state and have to be determined for each state individually. 
This property demonstrates one of the advantages of the DRN over QDs, where the possible changes of the couplings are 
orders of magnitude smaller.

We discuss charge transport through a DRN in the Coulomb blockade
regime near the $N = 0\leftrightarrow 1$ transition, i.e., when only one electron at a time can tunnel through 
a DRN between the source (S) and drain (D) electrodes. Throughout this paper 
we assume the magnetic field $B = 0$ and therefore neglect the electron spin. 
We show that one can tune the device parameters  so that it can work as:
({\em i)} single electron transistor \cite{kastner_1992,Janus} and ({\em ii}) electrical current rectifier. 
The paper is organized as follows: In Sec. \ref{sec2} we present a general theoretical background that will be needed to 
study the transport properties of DRNs. In  Sec. \ref{sec3} we demonstrate how by changing the parameters of
 the confinement potential we can control single electron tunneling and build a single electron transistor (SET). 
In Sec. \ref{sec4} we demonstrate that a DRN can be used as a current rectifier. The results are summarized in Sec.~\ref{sec6}.

\section{Basic formulas and mechanisms}\label{sec2}

We consider a quasi--two--dimensional circularly symmetric dot-ring nanostructure. 
The DRN, placed in the $xy$ plane, is defined by a specific confinement potential, 
that we will discuss below. We assume that the confinement in the $z$ (growth) direction is much 
stronger than the lateral confinement and consequently, the $xy$-plane motion and the vertical 
one can be decoupled. Then, we can write the electron wave function as a product
\begin{equation}
\psi(\bi{r})=\psi_{||}(\bi{r})\psi_z(z),
\end{equation}
where the vector $\bi{r}$ lies in the $xy$-plane. Additionally, we assume that the electron is always in the 
lowest energy state of a quantum well in $z$ direction and that the potential of the well in $z$ direction is 
infinite. With these assumptions $\psi_z(z)$ is given by
\begin{equation}
\psi_z(z)=\sqrt{\frac{2}{d}}\cos\left(\pi\frac{z}{d}\right),
\end{equation}
where $d$ is the height of the structure.

\begin{figure}[h]
\begin{center}
\includegraphics[width=0.67\textwidth]{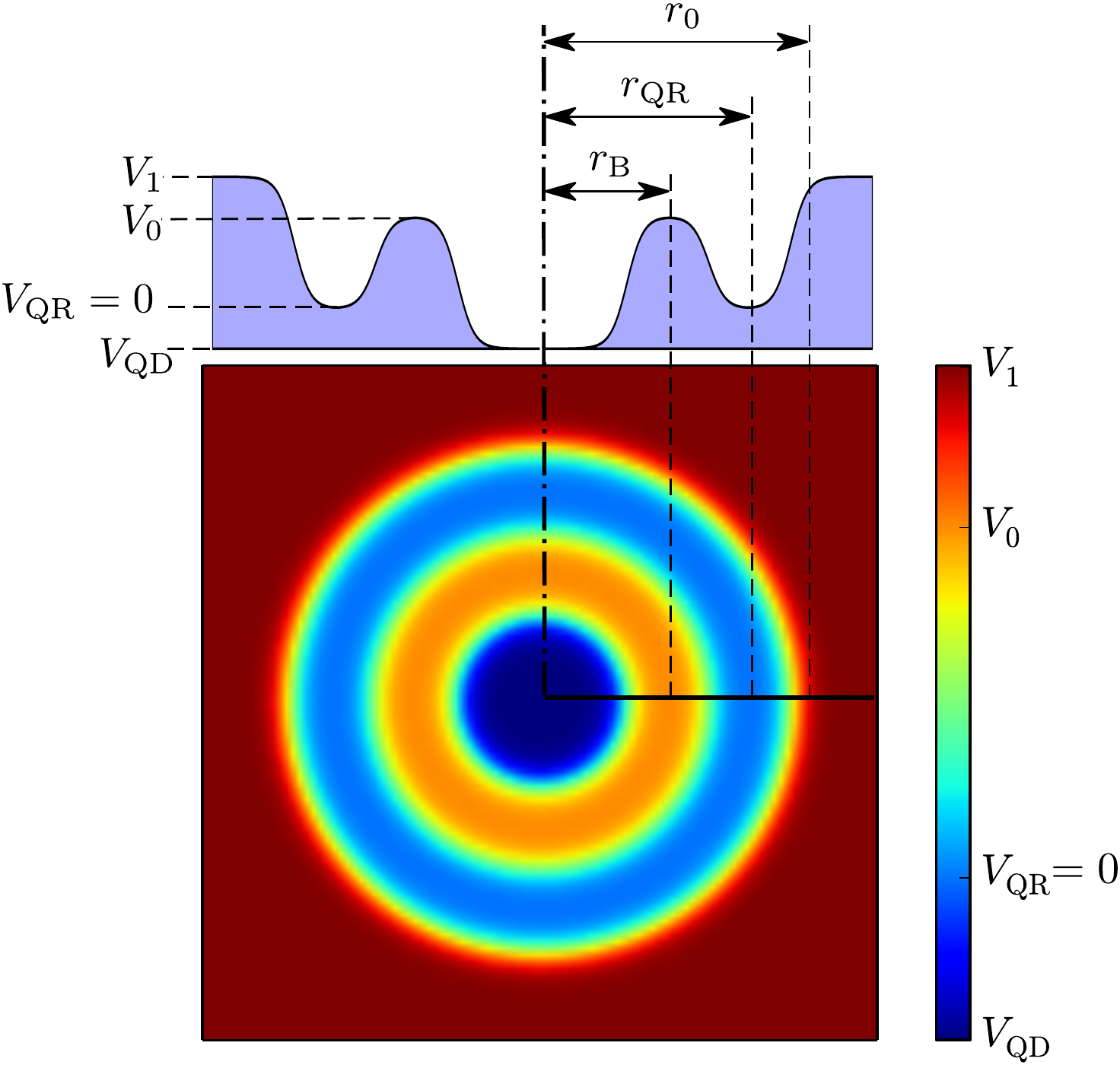}
\caption{The cross section and the top view of the potential forming the DRN with marked bottom of the QD potential 
($V_{\rm QD}$), bottom of the QR potential ($V_{\rm QR}$), top of the barrier potential ($V_0$) and the value of the
 potential outside the DRN ($V_1$). We assume the bottom of the quantum ring part as a reference value ($V_{\rm QR}=0$). 
This figure shows a DRN with the bottom of the QD below the bottom of the QR ($V_{\rm QD}<0$), but also the opposite 
situation ($V_{\rm QD}>0$) is possible.}
\label{crosssection}
\end{center}
\end{figure}

In order to discuss the in--plane confinement potential that forms the DRN we introduce the 
following notation 
 $|\bi{r}|=|(x,y)|=r$. Then the DRN is defined by a potential
 $V(r)$ and occupied by a {\it single} electron. The DRN is composed of a QD surrounded by a QR and separated 
from the ring by a potential barrier $V_0(r)$.
 A cross section and a top view of a DRN with explanation of symbols used throughout the text is presented 
in figure~\ref{crosssection}.

In particular, we assume in our model calculations the radius of the DRN $r_0=70$ nm. The depth of the quantum well 
forming the DRN is $V_1=90$ meV and the zero potential energy is set at the level of $V_{\rm QR}$, i.e., the potential 
well offset is equal $V_{\rm QD}$. The calculations are performed for InGaAs systems (with the effective electron mass 
$m^*= 0.067m_e$). The results are presented for $V_0=20$ meV and for the sample thickness $d=5$ nm, if not stated otherwise. 

\begin{figure}[h]
\begin{center}
\includegraphics[width=0.6\linewidth]{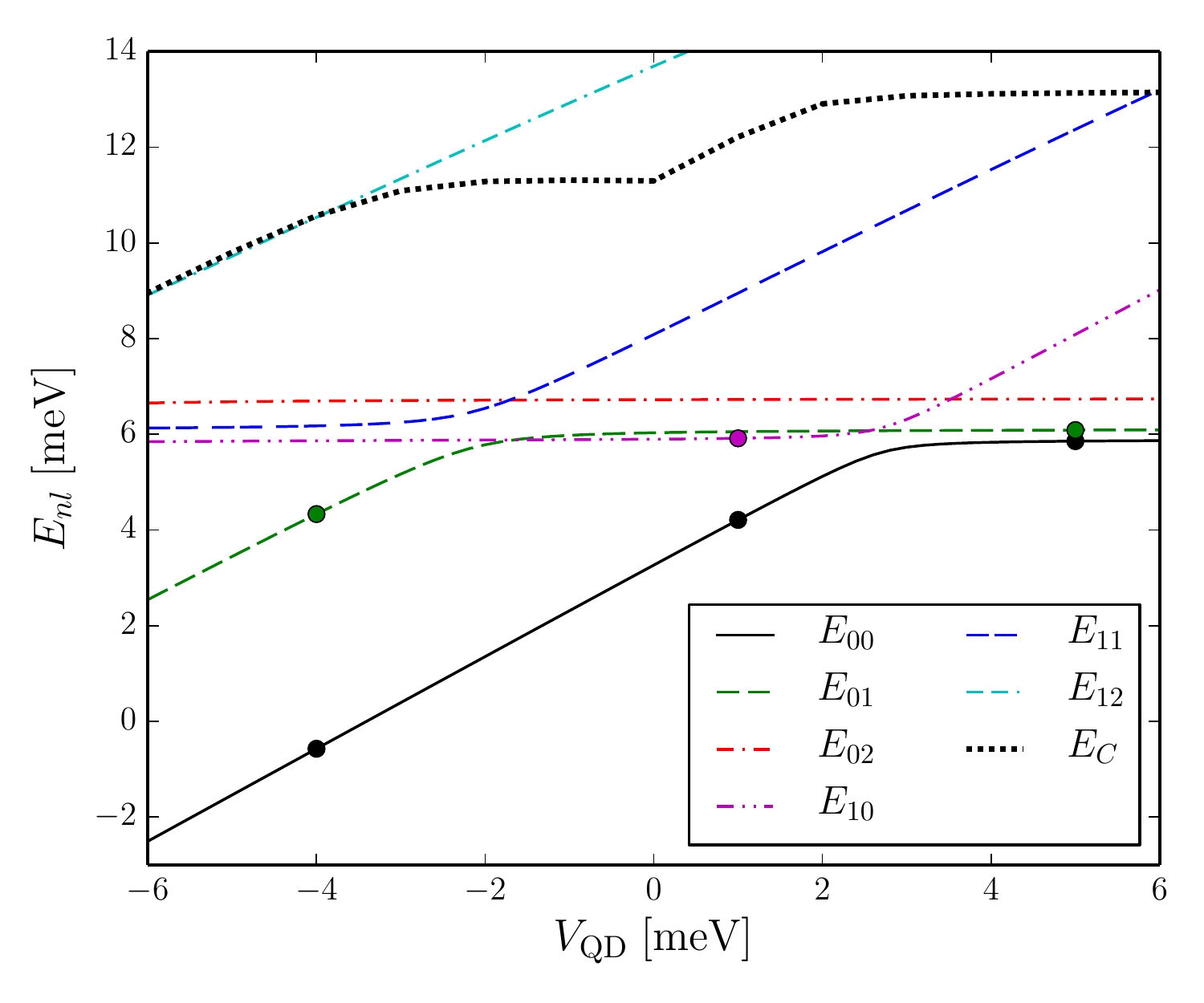}
\caption{Energy spectrum of a DRN as a function of the position of the bottom of the QD potential $V_{\rm QD}$.
The circles indicate states for which wave functions are presented in figure \ref{wavefunctions}.
The dotted line shows the energy of the lowest two--electron state (see text in Sec. 3).
}
\label{fig_levels}
\end{center}
\end{figure}

 We solve numerically the Schr\"odinger equation assuming the Gaussian form of $V(r)$
\cite{zipper}.
The energy spectrum consists of a set of discrete states $E_{nl}$ due to radial motion with radial quantum 
numbers $n=0,1,2,\ldots$, and rotational motion with angular momentum quantum numbers
 $l=0,\pm 1,\pm 2\ldots$. The energy spectrum as a function of $V_{\rm QD}$ is shown in 
figure \ref{fig_levels}. 
The states situated in QD exhibit an increase of  the energy with increasing $V_{\rm QD}$, whereas those 
situated in QR have the energy (nearly) constant.
The single particle wave function in the $xy$-plane, i.e., the DRN plane, is of the form 
\begin{equation}
\psi_{||}(\bi{r})\equiv \Psi_{nl}(r,\phi) = R_{nl}\left(r\right)\exp\left(i l \phi \right),
\label{eq_psi_nl}
\end{equation}
with the radial part $R_{nl}(r)$. 

\begin{figure}[h]
\begin{center}
\includegraphics[width=1.02\linewidth]{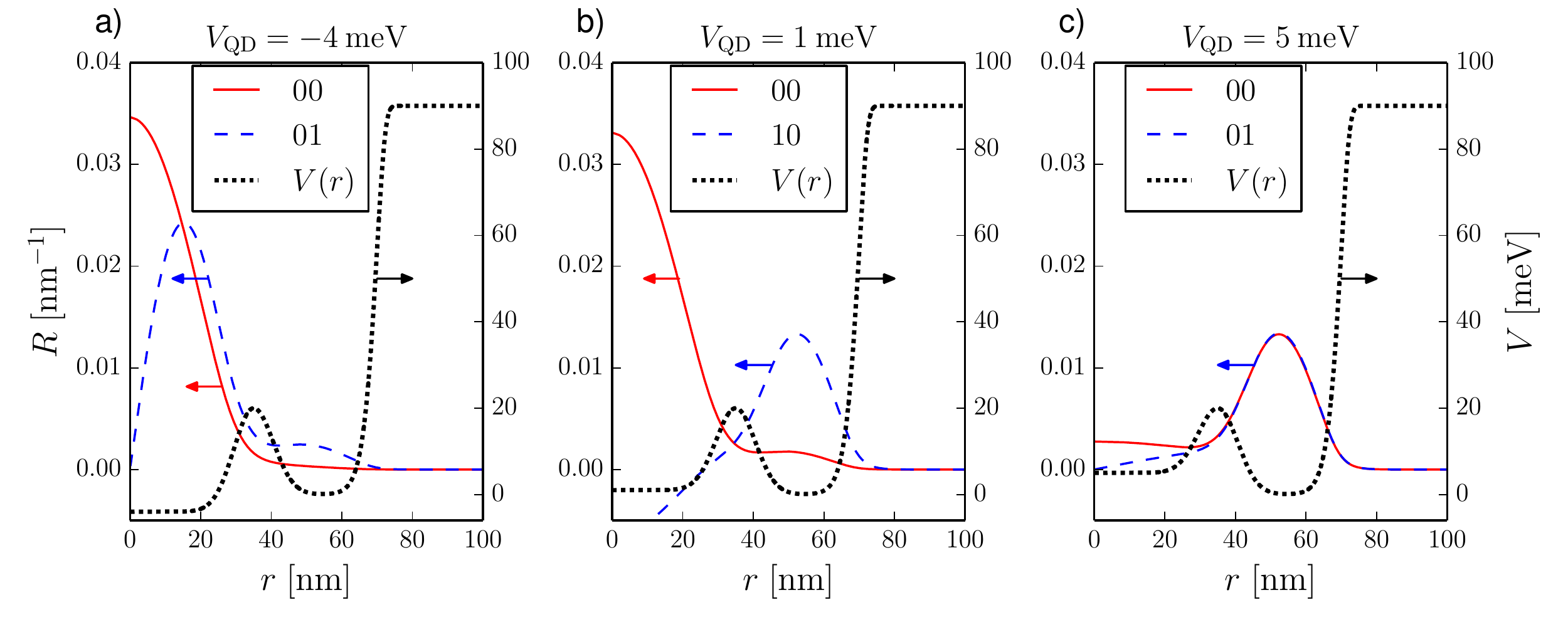}
\caption{Radial parts of the wave functions of the two lowest states (red and blue lines) for different 
values of $V_{\rm QD}$: -4 meV (a), 
1 meV (b), and 5 meV (c). The black dotted line shows the corresponding shape of the confining potential $V(r)$}.
\label{wavefunctions}
\end{center}
\end{figure}

As already mentioned, the main advantage of the DRN is the controllability of the shape and the distribution of the 
electron wave functions. 
For instance, if the minimum of the potential of the QD part $V_{\rm QD}$ is much deeper than the potential of the QR,
the electrons are located mainly in the QD and the effective size of the ground state (G) wave function is small.
On the other hand, if the ring's potential $V_{\rm QR}$ is much deeper  electrons occupy mostly states in the QR part 
and the G wave function is much broader. 
What is more, by fine--tuning the confinement potential we can control positions of individual
states. This way we are able to have, e.g., the ground state located in the QD, whereas the first excited state (E)
in the QR and so on. The distributions of the wave functions of the two lowest energy states for three
different values of $V_{\rm QD}$ are presented in figure \ref{wavefunctions}. One can see there the case where both 
the G and E wave functions are in the QD for $V_{\rm QD}=-4$ meV (figure \ref{wavefunctions}a), the E wave function 
is in the QR, whereas the G wave function is still in the QD for $V_{\rm QD}=1$ meV (figure \ref{wavefunctions}b), 
and finally, for $V_{\rm QD}=5$ meV both the G and E wave functions are in the QR (figure \ref{wavefunctions}c).

The DRN is coupled via tunnel barriers to the S and D electrodes.
We assume that one or a few ($n_0$) single--electron states are in the bias window
\begin{equation}
\mu_S > \epsilon_i >\mu_D,\ i=0,\ldots,n_0-1,
\end{equation}
where $\mu_S$ and $\mu_D$ are the respective chemical potentials, $i$ represents a set of 
quantum numbers $(n,l)$ and we start the numbering of the energy levels from $i=0\ (n=0,\:l=0)$ (the ground state).
Charge transport through a DRN depends crucially on the coupling strength 
of its states to the electrodes what is dependent on the wave function overlap that enters the electron tunneling
matrix element.
Because we are able to control the shape and distribution of the DRN wave functions, we can control the overlap,
 which, in turn, allows us to control the transport properties. 

The tunneling rates $\Gamma$'s are calculated microscopically for each state independently. We follow 
Bardeen's approach \cite{bardeen}, where 
\begin{equation}
\Gamma(\epsilon)=2\pi\sum_{\bm k}|t_{\bm k}|^2\delta\left(\epsilon-\epsilon_{\bm k}\right).
\end{equation}
 Within the framework of this method two separate sets of states are considered: one solves the Schr\"odinger 
equation for the DRN and one for the electrode. It is implicitly assumed that the wave functions of the two 
subsystems are orthogonal. 
\begin{figure}[h]
\begin{center}
\includegraphics[width=0.3\linewidth]{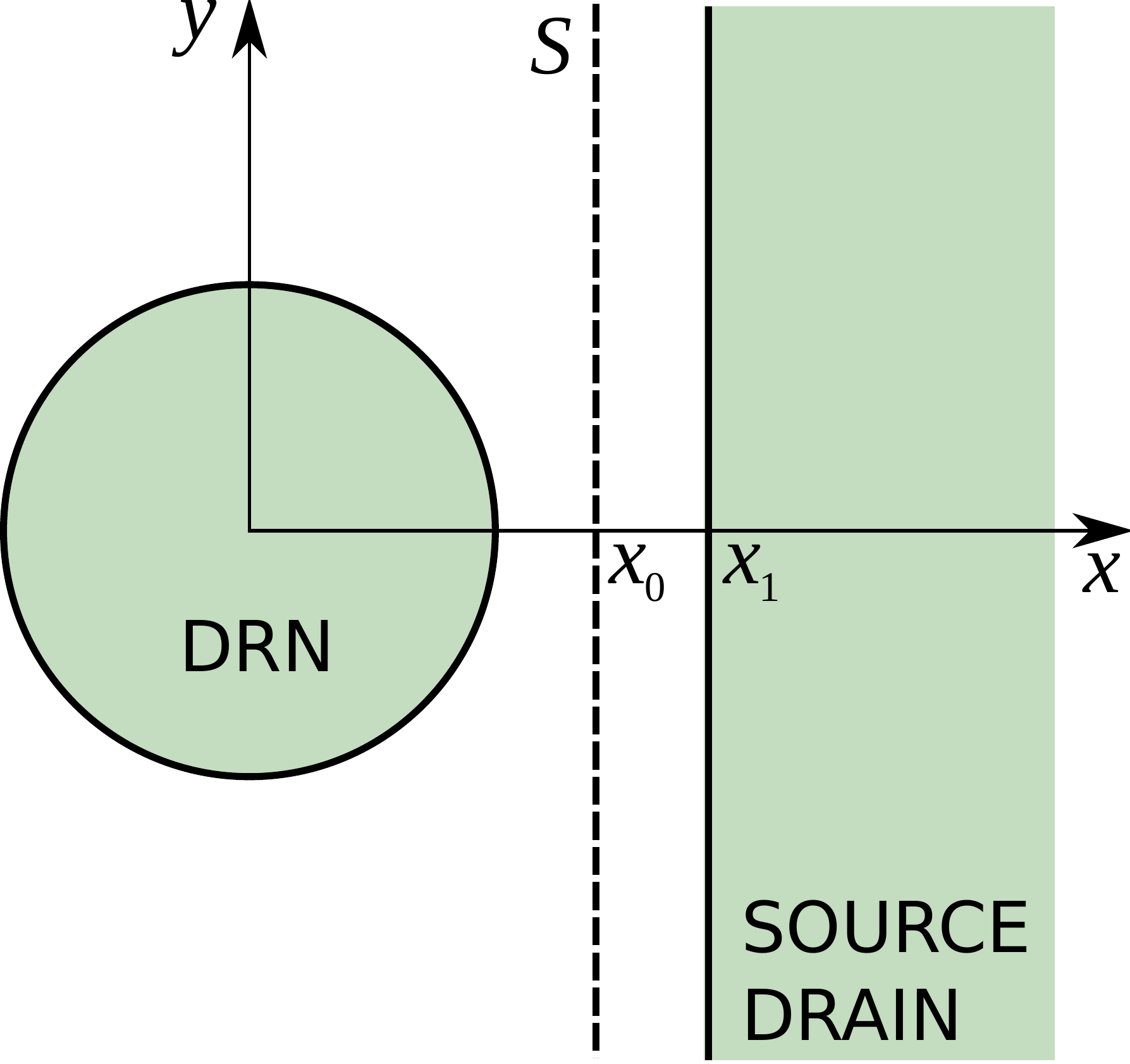}
\caption{The geometry of the setup used to calculate the couplings $\Gamma$ between the DRN and the electrodes.}
\label{coupling}
\end{center}
\end{figure}
Enforcing this assumption the Bardeen tunneling matrix element in two dimensions is given by 
\begin{equation}
t_{\bm k}=\frac{\hbar}{2m^*}\int^\infty_{-\infty}dy\left[\psi_{\bm k}^*({\bm r})\frac{\partial \Psi_{nl}({\bm r})}{\partial x}-\Psi_{nl}({\bm r})\frac{\partial \psi_{\bm k}^*({\bm r})}{\partial x}\right]_{x=x_0},
\end{equation}
where $\psi_{\bm k}({\bm r})$ are wave functions in the electrode and $\Psi_{nl}({\bm r})$ are wave functions in the DRN. The integral is calculated along the
 line $x=x_0$ (see figure \ref{coupling}). The explicit form of $t_{\bm k}$ is given in the Appendix.
%

\subsection{Phonon relaxation in a DRN}\label{sec_phonon}
When more that one state is included in the transport process one has to take into account mechanisms that allow transitions between the states, i.e., 
relaxation to the lower energy states. One of the most important and unavoidable mechanisms of scattering in solid state systems is the interaction with lattice phonons. 
For low-dimensional semiconducting systems as QDs and QRs the dominant process is the electron-acoustic phonon interaction.
This is because the energy distance between the 
electronic states is small comparing to the optical phonon energy \cite{bockelmann,piacente,stano}.

\begin{figure}[h]
\begin{center}
\includegraphics[width=0.6\linewidth]{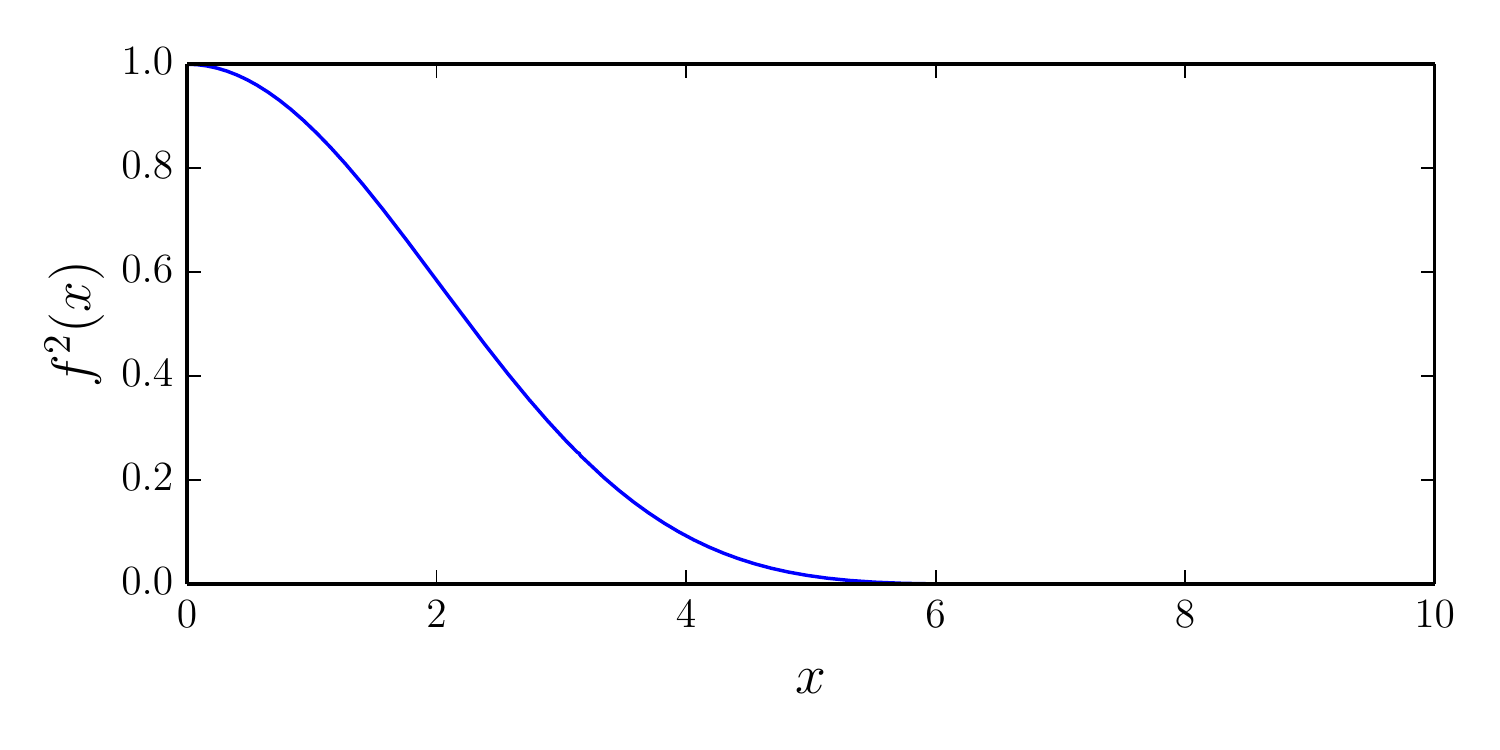}
\caption{Square of the function $f(x)$ given by equation (\ref{f}).}
\label{f_funct}
\end{center}
\end{figure}

In this section we calculate phonon emission relaxation rates due to interaction of an electron with piezoelectric (PZ) and deformational (DF) phonons. 
The electron-phonon scattering rate due to transition from the initial state $\psi^i(\bi{r})$ to the final state $\psi^f(\bi{r})$ with emission 
of an acoustic phonon can be calculated using Fermi's golden rule 
($w_{i\to f}\equiv\tau_{i\to f}^{-1}$), 
\begin{equation}
w_{i\rightarrow f} = \dfrac{2\pi}{\hbar} \sum_{\bi{q},\lambda}|\langle \psi^f|\bi{W}_{\lambda}(\bi{q})|\psi^i\rangle|^2 \delta(E_f-E_i-\hbar \omega_{q}),
\label{fermi_gr}
\end{equation} 
where $\bi{q}$ is the phonon wave vector, $E_f$ ($E_i$) is the energy of the final (initial) electron state, $\hbar \omega_q$ is the energy of a phonon and $\lambda$ is the polarization index. The interaction operator $\bi{W}_{\lambda}(\bi{q})$ is given by
\begin{equation}
\bi{W}_{\lambda}(\bi{q})=\bi{\Lambda}_{\lambda}(\bi{q})e^{-i \bi{q}\cdot\bi{r}},
\end{equation}
where $\bi{\Lambda}_{\lambda}(\bi{q})$ is the total scattering matrix element \cite{piacente}
\begin{equation}
|\bi{\Lambda}_{\lambda}(\bi{q})|^2 = |\bi{\Lambda}_{\rm LA}^{\rm DF}|^2+\sum_{\lambda={\rm LA,TA}}|\bi{\Lambda}_{\lambda}^{\rm PZ}(\bi{q})|^2.
\end{equation}
Because in this paper we focus mainly on transport properties of the DRN, we follow the reference \cite{piacente} 
and use angular averaged piezoelectric coupling matrix element for longitudinal and transverse phonon modes. Thus, 
the total electron-phonon scattering matrix element can be written as \cite{piacente,oliveira}
\begin{equation}
|\Lambda(\bi{q})|^2=\dfrac{\hbar}{2\rho c V|\bi{q}|}\left(D^2|\bi{q}|^2+P \right),
\end{equation}
where $D$ ($P$) is the deformational (piezoelectric) potential constant, $\rho$ is the crystal density, 
$c$ is the sound velocity, and $V$ is the volume. After Refs. \cite{piacente,oliveira} we assume 
$D=2.2\times 10^{-18}$~J
and $P=5.4\times 10^{-20}$~J$^2$m$^{-2}$. The total relaxation rate from state $n'l'$ 
to state $nl$ at $T=0$ can be written as
\begin{eqnarray}
w_{n'l'\rightarrow nl}&=&\dfrac{1}{4\pi^2 \hbar \rho c^2 } q_0^3\left(D^2q_0^2+P \right)\int_0^{2\pi}d\phi\int_0^{\pi/2}d\theta\sin\theta\nonumber\\
&\times &
\left\{ \left|\int_0^\infty dr \int_0^{2\pi}d\phi' e^{i(l-l')\phi'}e^{iq_0r\sin\theta\cos(\phi-\phi')}
r\:R_{nl}(r)R_{n'l'}(r)\right|^2\right. \nonumber\\
&\times&\left.
\left|\displaystyle\frac{2}{d}\int^{d/2}_{-d/2}\cos^2\left(\pi\frac{z}{d}\right)e^{-iz\:q_0\cos\theta} dz\right|^2\right\},
\label{relax_rate}
\end{eqnarray}
where $q_0=(E_{n'l'}-E_{nl})/\hbar c$ and $R_{nl}(r)$ is the radial part of the in-plane electron wave function 
$\psi_{||}(r)$. The integral over $z$ is given by $f\left(\frac{1}{2}d\,q_0\cos\theta\right)$, where
\begin{equation}
f(x)\equiv\frac{\pi^2}{\pi^2 x-x^3}\sin x
\label{f}
\end{equation}
and the integral over $\phi'$ can be expressed by the Bessel function. Finally, the relaxation rate can 
be written as
\begin{equation}
w_{n'l'\rightarrow nl}=\dfrac{2\pi}{\hbar \rho c^2} q_0^3\left(D^2q_0^2+P \right)
\int_0^{\pi/2}d\theta\sin\theta\: F^2_{n'l',nl}(\theta),
\end{equation}
where
\begin{equation}
F_{n'l',nl}(\theta)=f\left(\frac{1}{2}d\,q_0\cos\theta\right)
\int_0^\infty dr J_{|l-l'|}(q_0r\sin\theta)r\:R_{nl}(r)R_{n'l'}(r).\label{FT}
\end{equation}
Apart from material constants the factors that affect this rate are the energy gap between states 
$n'l'$ and $nl$, i.e., $E_{n'l'}-E_{nl}$, the mutual distribution of the wave functions given by $R_{nl}(r)$ 
and $R_{n'l'}(r)$, and the thickness of the structure $d$. 
Function $F_{n'l',nl}(\theta)$ describes how
phonons emitted in different directions contribute to the relaxation process. The relaxation through
phonons emitted at a given angle $\theta$ depends on $d$ through function $f$ given in equation (\ref{f}). 
On the one hand, since its argument is $\frac{1}{2}d\,q_0\cos\theta$ the relaxation is independent 
of the thickness $d$ for phonons with wave vectors parallel to the $xy$ plane ($\theta=\pi/2$). On 
the other hand, there is a strong dependence of the relaxation rate on $d$ for phonons emitted in the 
direction perpendicular to nanostructure. Figure \ref{f_funct} shows $f^2(x)$. One can see there that for
$\theta=0$ a significant contribution to the relaxation comes only from phonons with wavelength larger then $d$.

\begin{figure}[h]
\begin{center}
\includegraphics[width=0.9\linewidth]{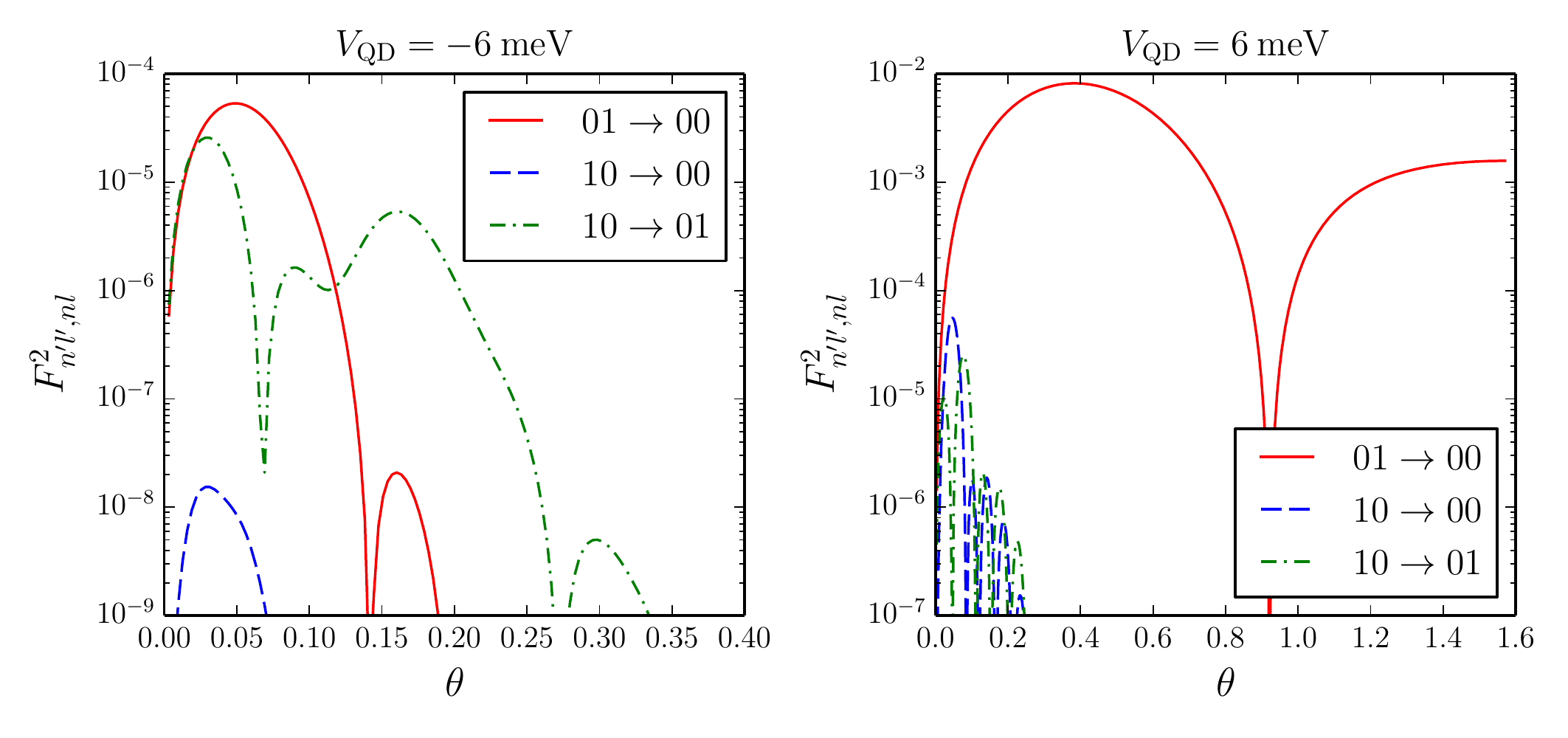}
\caption{Square of the function $F_{n'l',nl}(\theta)$ given by equation (\ref{FT}) for two different 
values of $V_{\rm QD}$ for a DRN with $d=5\:{\rm nm}$.}
\label{f_theta}
\end{center}
\end{figure}

Figure \ref{f_theta} shows the square of the function $F$ given by equation (\ref{FT}) for $d=5\:{\rm nm}$. 
The value of $q_0$ and the
radial parts of the wave functions $R_{nl}(r)$ and $R_{n'l'}(r)$ have been calculated for two different
 values of $V_{\rm QD}$. 
In the case presented in the left panel ($V_{\rm QD}=-6\:{\rm meV}$) the bottom of the QD potential is much
 below the 
bottom of the QR potential. Therefore, the wave functions are situated mainly in the QD part of the DRN, 
where the energy 
level spacing is large (see figure \ref{fig_levels}). On the other hand, the right panel presents the case of 
$V_{\rm QD}=6\:{\rm meV}$ where the bottom of the QD potential is above the
bottom of the QR potential. In this situation the wave functions are situated mainly in the QR part of
 the DRN and, as it infers from figure  \ref{fig_levels}, the level spacing is small. 
Figure \ref{f_theta} allows one to analyze the directions of phonons emission in two
presented cases. Namely, in the left panel ($V_{\rm QD}=-6\:{\rm meV}$) function
$F^2_{n'l',nl}(\theta)$ describing 
the phonon emission is peaked within the range of low values of $\theta$. This indicates that 
phonons are emitted mostly perpendicularly to the DRN, which is the direction of the strongest confinement
\cite{bockelmann}. In the complementary case presented in the right panel ($V_{\rm QD}=6\:{\rm meV}$), 
it is not possible to point out a specific direction of emission (it varies with a particular transition 
$n'l' \rightarrow nl$, note the difference in the scales on the horizontal axes).

\begin{figure}[h]
\begin{center}
\includegraphics[width=0.6\linewidth]{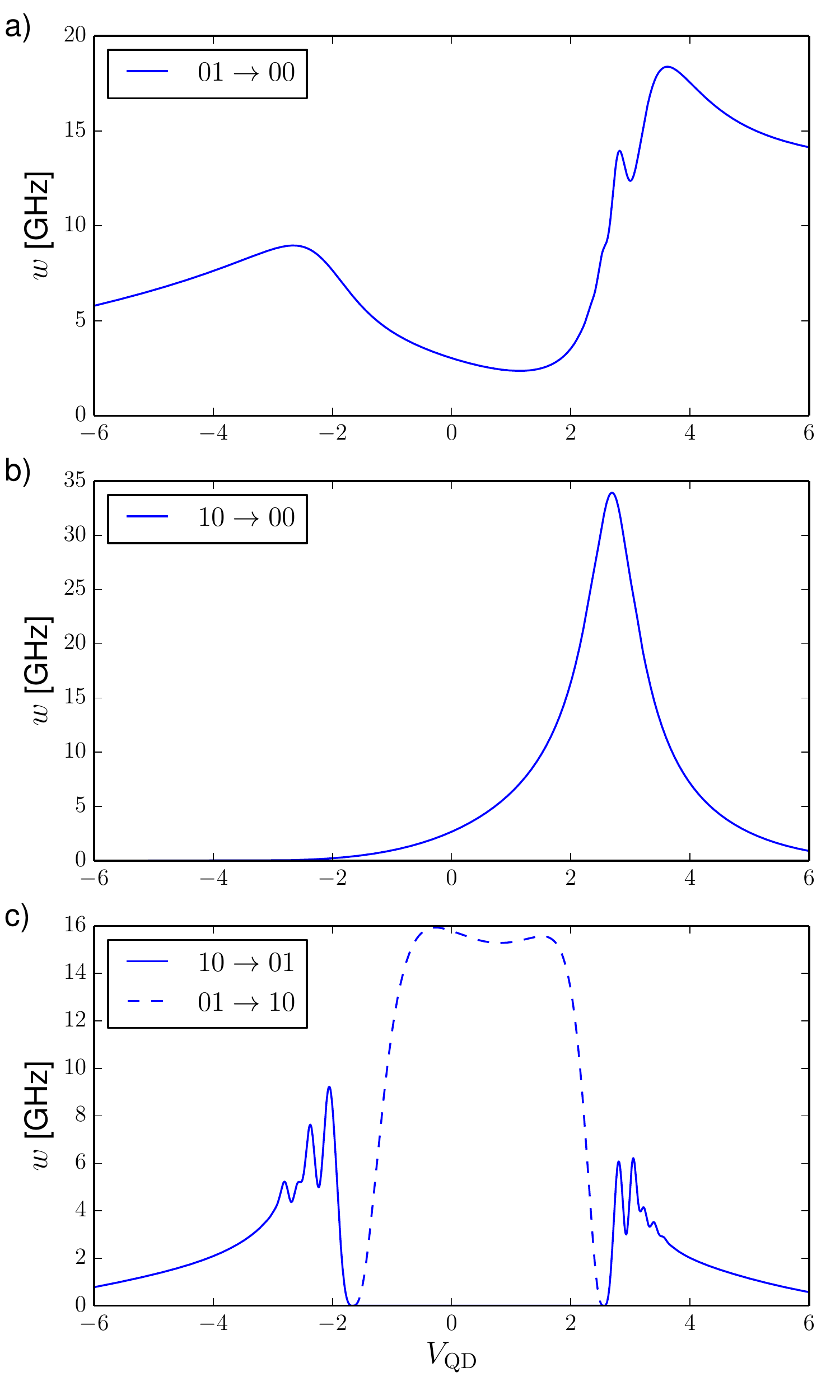}
\caption{Illustrations of the dependence of different relaxation rates on $V_{\rm QD}$ for $d=5$~nm. 
See the text for an explanation.}
\label{relax_rates}
\end{center}
\end{figure}

Figure \ref{relax_rates} shows typical examples of how the relaxation rates are affected 
by changing the position of the bottom of the QD potential $V_{\rm QD}$ from below to above the bottom of the QR 
potential ($V_{\rm QR} =0$). 
Figure \ref{relax_rates}a shows the relaxation rate from state $(n=0,\:l=1)$ to the ground state $(n=0,\:l=0)$. 
For large, negative value of $V_{\rm QD}$ both the wave functions $\Psi_{00}$  and $\Psi_{01}$ are positioned in the 
QD what gives fast relaxation with rates $w$ of the order of GHz. With increasing $V_{\rm QD}$ the excited state moves 
over to the QR leading to the decrease of the overlap between the wave functions what, in turn, results in the 
decrease of $w$. For further increased $V_{\rm QD}$ both the ground and the excited state are situated in the QR, 
the overlap increases and the relaxation rate increases again. 
Figure \ref{relax_rates}b presents the relaxation rate from state $(n=1,\:l=0)$ to the ground
state. In this case for small $V_{\rm QD}$ the ground state is localized in the QD, while the excited state 
$(n=1,\:l=0)$ is mostly positioned in the QR. Therefore, the overlap of the corresponding wave functions is 
small, what in turn results in a slow relaxation. With the increase of $V_{\rm QD}$ the ground state starts to 
move over to the QR and the overlap with $\Psi_{10}$ starts to rise. One observes it as a sharp increase of 
the relaxation rate $w_{10\rightarrow 00}$. However, with further increase of  $V_{\rm QD}$
the ground state remains in the QR, yet the excited state moves over to the QD 
and changes sign. 
This altogether causes the decrease of the relaxation rate. 
Figure \ref{relax_rates}c presents the relaxation rate between states $(n=1,\:l=0)$ and $(n=0,\:l=1)$, a transition that
is a part of an indirect relaxation process.  What is clearly 
visible at first glance is the abrupt decrease of $w_{10\leftrightarrow 01}$ when the states cross 
(compare figure \ref{fig_levels}). In the range of $V_{\rm QD}$ from -2 meV to 2 meV the overlap between $\Psi_{10}$ 
and $\Psi_{01}$ is strong. Thus, the resulting values of the relaxation rate are high.

\begin{figure}
\begin{center}
\includegraphics[width=0.6\linewidth]{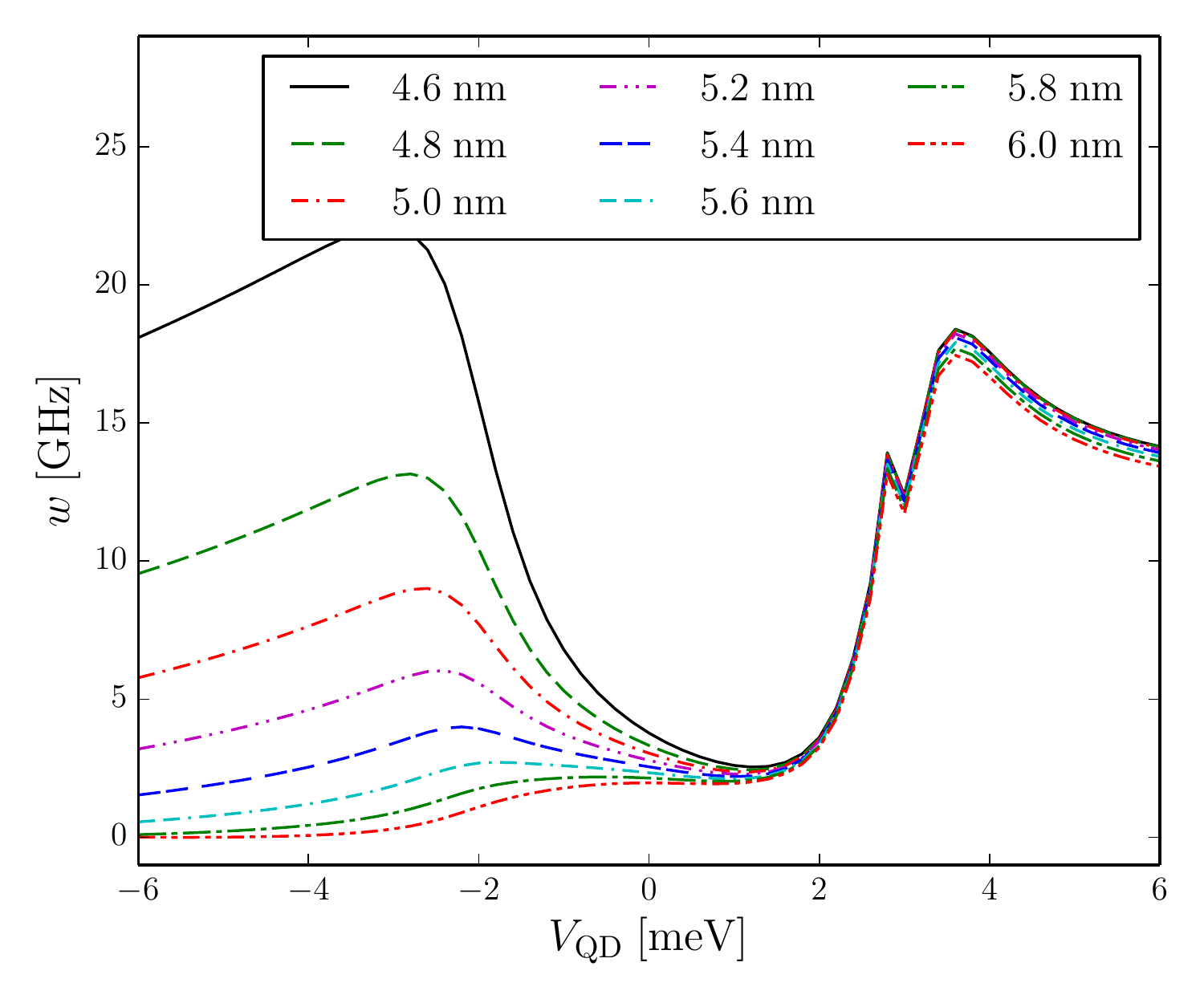}
\caption{Relaxation rate $w_{01\rightarrow 00}$ as a function of $V_{\rm QD}$ for different
values of sample thickness.}\label{w_vs_d}
\end{center}
\end{figure}

As it was already mentioned, the phonon emission at different angles is strongly dependent on the distribution
of the wave functions (see figure \ref{f_theta}). This, in turn affects the dependence of the relaxation rates
on the sample thickness $d$.
In figure \ref{w_vs_d} we present the relaxation rate $w_{01\rightarrow 00}$ as a function of $V_{\rm QD}$. Its 
dependence on $d$ is pronounced for low values of $V_{\rm QD}$,
when the wave functions are situated mainly in the QD region of the DRN. 
In accordance with what was reported in Ref. \cite{piacente} for QD's, we also observe oscillations of the 
relaxation rates as a function of the sample thickness $d$ in the range of negative values of $V_{\rm QD}$
(yet in figure \ref{w_vs_d} we present only a part of the noted oscillations). In the range of low values
of $V_{\rm QD}$ the level spacings are of the order of single meV's (figure \ref{fig_levels}). 
The corresponding wavelengths of emitted phonons are of the order of the sample thickness (couple of nm's).
This match results in the strong increase of the relaxation rates for such $V_{\rm QD}$. In the complementary
range of $V_{\rm QD}$, where the bottom of the QD potential is above the bottom of the QR potential, the level 
spacings are of order of 0.1 meV. Then, the corresponding wavelengths of emitted phonons are significantly 
larger than the sample thickness and we do not observe any strong dependence of $w_{01\rightarrow 00}$ on the  
thickness of the structure $d$. Additionally, we would like to stress the strong dependence of the relaxation
rates for all other transitions $w_{n'l'\rightarrow nl}$ on $d$ (not shown).
\section{Single electron transistor}\label{sec3}
We are now ready to discuss the transport properties of the DRN. We evaluate the current given the energy 
spectrum and the subsequent relaxation and tunnel rates. We discuss below sequential tunneling current in 
the Coulomb blockade regime near the $ N = 0 \leftrightarrow1$ transition and neglect higher order tunneling  
events \cite{R.Hanson}. 
In the pure quantum dot case SET is switched to the conducting state when some energy level, shifted 
by gate voltage, enters
the bias window. The current exhibits then the current peak \cite{hans}. In case of the DRN the mechanism is 
different. We keep one or a few states in the bias window and manipulate the distribution of the wave 
functions to get a proper transistor behavior. The principle of operations in is to control
the energy states and tunnel couplings by means of gate voltages $V_{\rm QD}$ and $V_0$.
Many investigations of transport behavior make use of the high tunability of the tunnel barriers by 
applying voltage pulses \cite{R.Hanson}. In our approach we utilize instead the high tunability of the 
electron states in the DRN while keeping the barrier parameters constant. We assume that the bias window 
is smaller than the charging energy $E_C$, so that only a single electron at a time can be transmitted 
through the DRN, $|\mu_S -\mu_D| < E_C$. In order to determine for what values of the
 model parameters this condition can be fulfilled we calculate $E_C$ for the DRN.
The interaction energy of two particles confined in the potential $V(r)$ was calculated using the configuration 
interaction approach \cite{shavitt_1977}, which is an exact diagonalization method for solving the nonrelativistic 
Schr\"{o}dinger equation for a multi-particle system. From the single-particle orbitals $\Psi_{nl}$ 
we constructed the basis of the Slater determinants $|S_{\mu} \rangle$. Then, the two-particle Hamiltonian
was diagonalized and the exact eigenstates were found where the $\nu$-th eigenfunction is in the form of 
the linear combination of the Slater determinants:
\begin{equation}
 \Phi_{\nu} = \sum_{\mu}c_{\mu \nu} |S_{\mu}\rangle.
\end{equation}
Coefficients $c_{\mu \nu}$ were calculated by the two--particle Hamiltonian diagonalization.
In figure \ref{fig_levels} the energy $E_C$ of the lowest two--particle state is shown by the black dotted line 
as a function of 
$V_{\rm QD}$. It can be seen there that for the analyzed range of $V_{\rm QD}$, $E_C$ is larger then the 
five lowest single--particle levels.

The current through the DRN is calculated with the help of the rate equations. Assuming that $n_0$ 
energy levels lie in the bias window $\mu_S-\mu_D$, the time evolution of the occupation probability 
$\rho_j$ of a given DRN state $j$ can be expressed by the following formula:
\begin{equation}
\dot{\rho}_j=\Gamma^S_j\left(1-\sum_{i=0}^{n_0-1}d_i\rho_i\right)+\sum_{i=j+1}^{n_0-1}d_iw_{i\rightarrow j}\rho_i-\left(\Gamma^D_j+\sum_{i=0}^{j-1}d_iw_{j\rightarrow i}\right)\rho_j.
\label{rate_eq}
\end{equation}
Indices $i$ and $j$ denote pairs of quantum numbers $(n,l)$ with $i=0$ describing the ground state $(n=0;\:l=0)$; 
the states are ordered so that $\epsilon_i>\epsilon_j$ if $i>j$ and
\begin{equation}
d_i=\begin{cases} 1 &\mbox{if } i\ \mbox{denotes a state with }l=0,\\
2 &\mbox{if } i\ \mbox{denotes a state with }l>0
\end{cases}
\end{equation}
is the orbital degeneracy of $i$--th state ($E_{nl}=E_{n,-l}$). 
\begin{figure}[h]
\begin{center}
\includegraphics[width=0.7\linewidth]{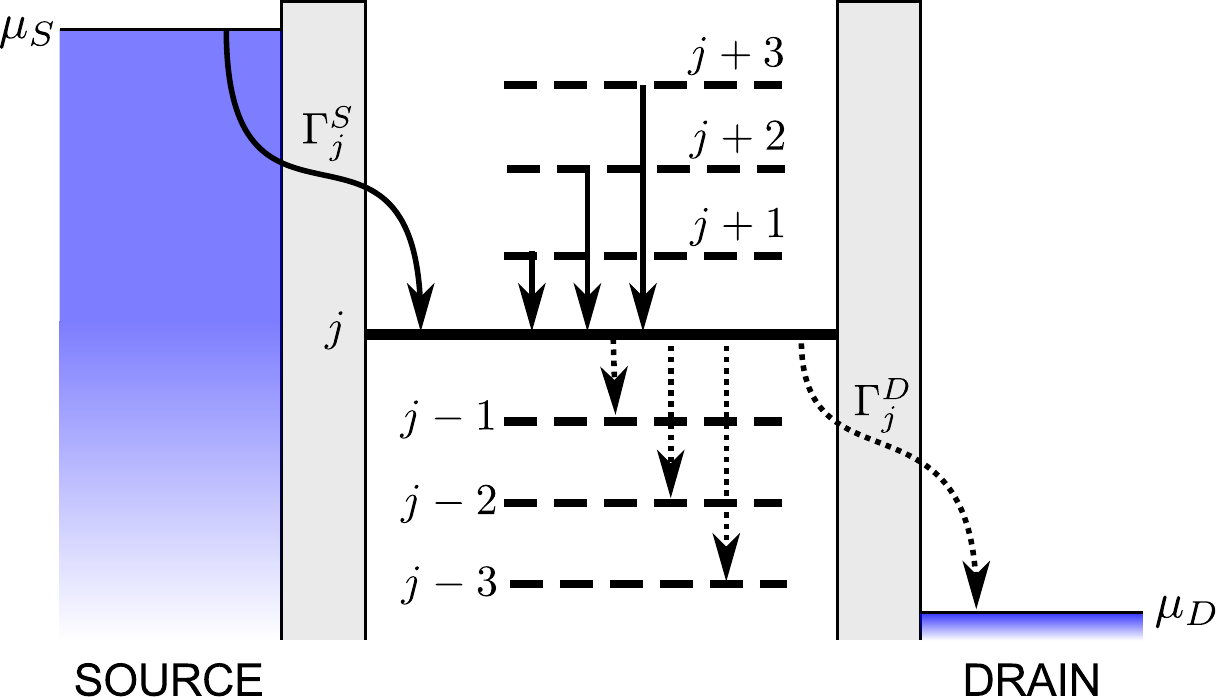}
\caption{Processes included in equation \ref{rate_eq}. The solid (dotted) S--shape line represents 
the transfer of an electron  from the source electrode to state $j$ (from state $j$ to the drain electrode). 
The vertical solid (dotted) lines represent relaxation from states above state $j$ to state $j$ (from state $j$ 
to states below state $j$).}
\label{relax_levels}
\end{center}
\end{figure}
The processes that enter equation (\ref{rate_eq}) are illustrated in figure \ref{relax_levels}. The first term on the 
r.h.s. describes the rate at which electrons tunnel from the S electrode. Due to the Coulomb blockade such a 
transfer is possible only when none of the DRN states is already occupied, what is ensured by the expression 
in parentheses. The second term describes relaxation to state $j$ from higher--lying states. The last term 
describes tunneling to the D electrode and relaxations from state $j$ to lower--lying states. If $j=n_0$ ($j=0$)
the second (third) sum in equation (\ref{rate_eq}) should be omitted since there are no states from (to) which 
relaxation is possible.

As we are interested in the steady--state current we put $\dot{\rho}_j=0$ in equation (\ref{rate_eq}). Then, 
introducing $\bar{\rho}_i\equiv d_i \rho_i$, the system of equations for $\rho_j$ can be rewritten as
\begin{eqnarray}
\displaystyle
\hspace*{-50pt}\left(\begin{array}{ccccc}
\Gamma^S_0+\displaystyle\frac{1}{d_0}\Gamma^D_0 & \Gamma^S_0-w_{1\rightarrow 0} & \Gamma^S_0-w_{2\rightarrow 0} & \ldots \\ 
\Gamma^S_1 & \Gamma^S_1+\displaystyle\frac{1}{d_1}\left(\Gamma^D_1+w_{1\rightarrow 0}\right) & \Gamma^S_1-w_{2\rightarrow 1} & \ldots \\ 
\Gamma^S_2 & \Gamma^S_2 & \Gamma^S_2+\displaystyle\frac{1}{d_2}\left(\Gamma^D_2+w_{2\rightarrow 0}+w_{2\rightarrow 1}\right) & \ldots \\
\Gamma^S_3 & \Gamma^S_3 & \Gamma^S_3 & \ldots \\ 
\cdots & \cdots & \cdots & \ldots \\ 
\Gamma^S_{n_0-1} & \Gamma^S_{n_0-1} & \Gamma^S_{n_0-1} & \ldots  
\end{array}\right. \nonumber \\
\nonumber \\
\nonumber \\
\hspace*{-50pt}\left.\begin{array}{cc}
\ldots & \Gamma^S_0-w_{(n_0-1) \rightarrow 0} \\
\ldots & \Gamma^S_1-w_{(n_0-1) \rightarrow 0} \\
\ldots & \Gamma^S_2-w_{(n_0-1) \rightarrow 0} \\
\ldots & \Gamma^S_3-w_{(n_0-1) \rightarrow 0} \\
\ldots & \ldots\\
\ldots & \Gamma^S_{n_0-1}+\displaystyle\frac{1}{d_{n_0-1}}\left(\Gamma^D_{n_0-1}+\sum_{i=0}^{n_0-2} w_{(n_0-1)\rightarrow i}\right)
\end{array}\right)
\left(\begin{array}{c}
\bar{\rho}_0 \\ \bar{\rho}_1 \\ \bar{\rho}_2 \\ \bar{\rho}_3\\ \cdots \\ \bar{\rho}_{n_0-1}\end{array}\right)=
\left(\begin{array}{c}\Gamma^S_0 \\ \Gamma^S_1 \\ \Gamma^S_2 \\ \Gamma^S_3 \\ \cdots \\ \Gamma^S_{n_0-1}\end{array}\right)
\label{matrix_eq}
\end{eqnarray}
In the absence of magnetic field $d_0$ is always equal to 1, but we have left it in the first row for consistency of notation.

The solutions can be easily found in limiting cases of a very slow relaxation and of a very weak couplings to the electrodes. In the 
former case, when we put $w_{i\rightarrow j}\rightarrow 0$ for all $i$ and $j$, the solution for a symmetric coupling ($\Gamma^S_i=\Gamma^D_i$) 
takes on a simple form $\rho_i=1/(n_0+1)$. In the 
latter case, when we put $\Gamma^S_i,\Gamma^D_i\rightarrow 0$ for all $i$, we get $\rho_i= 0$ for all $i>0$, what
means that only the ground state participates in the transport.
In a general case the system of linear equations (\ref{matrix_eq}) can be solved analytically, but with increasing 
$n_0$ the formulas quickly become very long. With the help of the occupation probabilities $\rho_i$, the 
steady--state current can be expressed as a sum of currents $I^{\rm in}_i$ carried by electrons which tunnel from the S electrode to all the states that lie in the bias window. Since in the steady state the 
currents through both the barriers are equal, the total current can be also expressed as a sum of currents $I^{\rm out}_i$ carried by 
electrons which tunnel from the DRN to the D electrode. 
Then, the total current can be written as
\begin{equation}
I=\sum_{i=0}^{n_0-1}I^{\rm in}_i=\sum_{i=0}^{n_0-1}I^{\rm out}_i,
\label{tot_current}
\end{equation}
where the currents carried by electrons tunneling to and from individual levels are given by
\begin{equation}
I^{\rm in}_i=e\Gamma^S_i(1-\rho_i),\ \ I^{\rm out}_i=e\Gamma^D_i\rho_i.
\label{current_i}
\end{equation}

We assume, following the experiments in Refs. \cite{Fudzi,enslin}, $k_BT\approx 0.01\:$ meV. 
Thus, in our studies both the energy spacings and the bias window are much larger than the thermal 
energy  and we neglect the temperature smearing out of the energy levels. Then for a forward bias 
($\mu_S-\mu_D>0$)
the electrons can tunnel only from the S electrode to the DRN and then from the DRN to the D electrode.
The tunnel rates in equation (\ref{matrix_eq}) depend on the DRN geometry. We assume it so that $\Gamma$ for 
the most strongly coupled state is equal to 5 GHz and calculate all the other couplings accordingly.
That way all the tunnel rates are in the range of the experimentally accessible values \cite{Fudzi1}. 

\begin{figure}[h]
\begin{center}
\includegraphics[width=\linewidth]{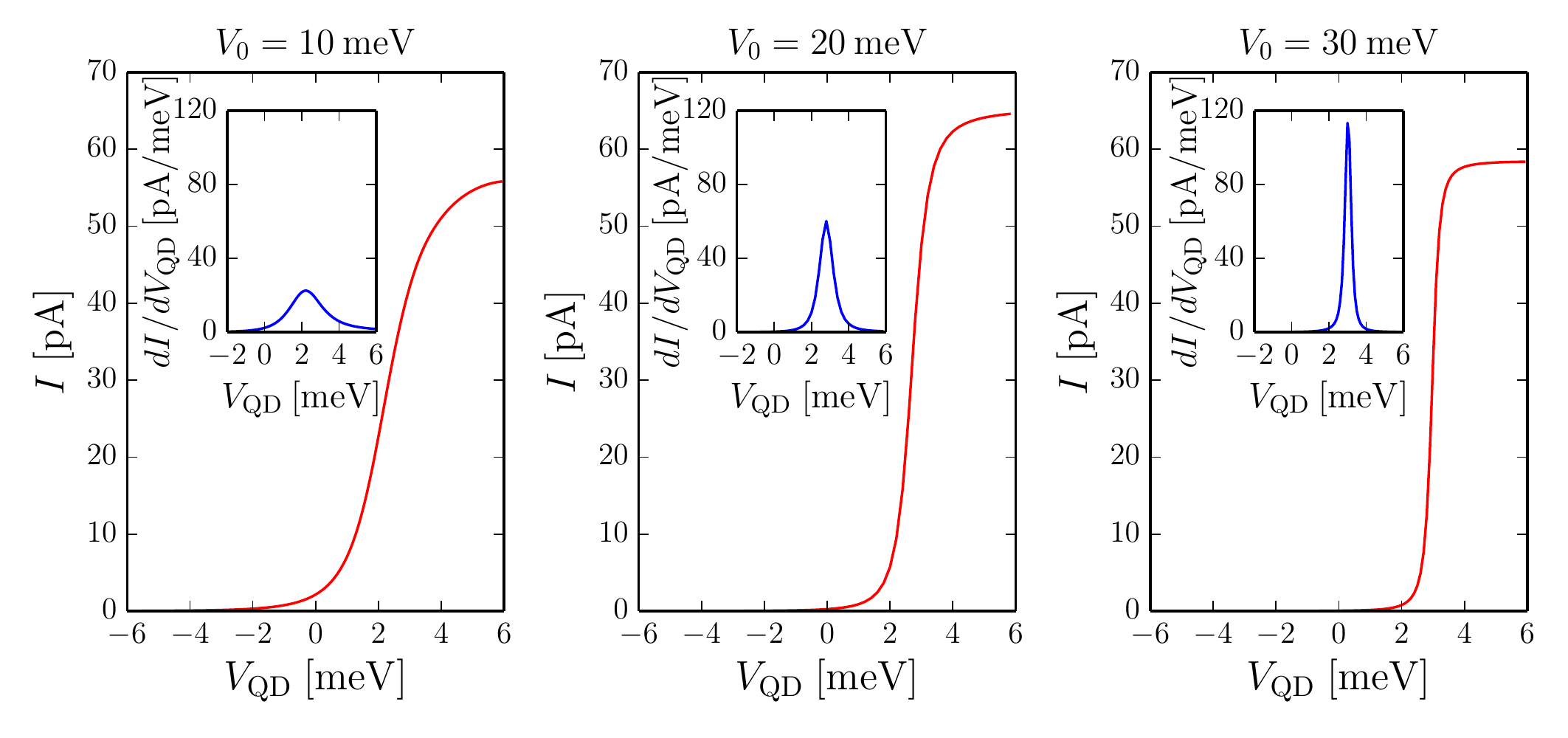}
\caption{The current as a function of $V_{\rm QD}$ for three values of $V_0$: 10, 20 and 30 meV.
$\Gamma=500$ MHz has been assumed in all the cases. 
}
\label{curr_vs_V0}
\end{center}
\end{figure}

We start with the low bias regime where only the ground state lies in the bias window ($n_0=1$). 
In this case the total current is given 
by equation (\ref{current_i}) with the requirement that $I^{\rm in}_0=I^{\rm out}_0$:
\begin{equation}
I=e\:\frac{\Gamma^S_G\Gamma^D_G}{\Gamma^S_G+\Gamma^D_G}.
\end{equation}
We assume symmetric tunneling barriers to the S and D electrodes $\Gamma_{i}^S = \Gamma_{i}^D$ and 
numerically calculate their values. The mechanism of the SET is as follows:
if the bottom of the QD potential is significantly below the bottom of the QR potential, the 
ground state is localized in the QD near the center of the DRN. Since for a deep enough QD potential
the wave function decreases almost exponentially with increasing distance from the QD, its overlap
with the electrode's wave functions is negligible. With an increase of $V_{\rm QD}$ this state moves 
to the outer (QR) part of the structure where it has much larger overlap with the states in the electrodes. 
This results in a strong increase of the current.
%
We calculate the current as a function of the gate voltage $V_{\rm QD}$ for three different values of $V_0$.
 Results  are presented in figure \ref{curr_vs_V0}.
One can see in this figure that the steepness of the $I(V_{\rm QD})$ characteristics increases significantly 
with an increase of $V_0$. If $V_0$ is small, the ground state wave function gradually moves with increasing
$V_{\rm QD}$ towards the outer part of the DRN. This leads to a slow increase of the current. On the other 
hand, if $V_0$ is large, there are two pronounced minima of the confining potential (one in the QD and and one 
in the QR) and with increasing $V_{\rm QD}$ at some point the ground state wave function ``jumps'' from the QD 
to the QR. It results in a step--like $I(V_{\rm QD})$ characteristics. This dependence is clearly seen in the 
insets in figure \ref{curr_vs_V0}, where the differential conductance $dI/dV_{\rm QD}$ is presented for different 
heights of the barrier $V_0$ separating the DRN.

\begin{figure}[h]
\begin{center}
\includegraphics[width=0.9\linewidth]{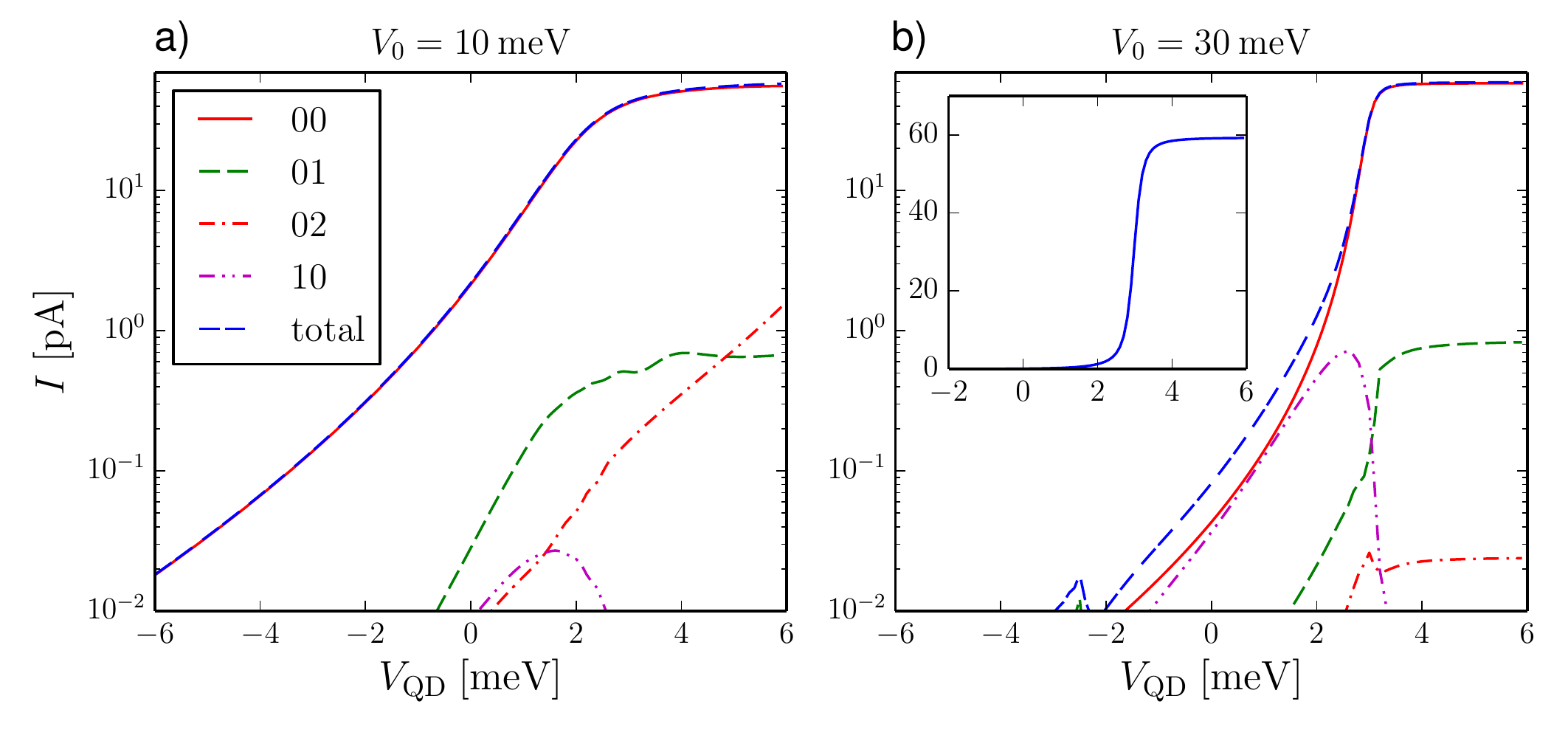}\\
\caption{The total current along with contributions that flow through particular DRN's levels for 
 $\Gamma=500$ MHz,  $V_0=10$ meV (a) and $V_0=30$ meV (b). 
In both panels the meaning of the lines is the same. The inset in the right panel shows the
total current in non--logarithmic scale.
}
\label{currents_4levels}
\end{center}
\end{figure}

The low bias limit, however, not always can be reached. In many cases the evolution 
of the DRN's spectrum shows several level crossings when $V_{\rm QD}$ is changed. In such cases the bias 
$\mu_S-\mu_D$ cannot be adjusted to include only
the ground state for all values of $V_{\rm QD}$ for which the transistor effect occurs. We see in figure \ref{fig_levels}
that depending on $V_{\rm QD}$ two or three (or even more) states have to be accounted for.
%
%
%
It is known that transport in the Coulomb blockade regime can be suppressed by the occupation of excited states \cite{Weis}.
Therefore, we analyze below how the system should be designed to get a good transistor behavior in the high bias regime.
The dc currents are calculated as steady state solutions to the rate equations 
[equations (\ref{matrix_eq}-\ref{current_i})] for the occupation probabilities 
for all relevant states considering all the processes transferring electrons, namely the tunnel rates $\Gamma$'s  and
 subsequent relaxation rates $w$'s. 
It turns out that the major factor that determines the current is the relations between the tunnel and 
relaxation rates. 
When the coupling to the electrodes is small compared to the corresponding relaxation rates, most of electrons that
travel through the DRN will relax to the ground state before leaving the DRN by tunneling to the drain electrode. Then,
the total current is determined mostly by the transport only through the ground state even if some of the excited states 
are in the bias window. In this case the system behaves like in the described above low bias limit. 
This situation is illustrated in figure \ref{currents_4levels}. 

What we have also noted from our numerical calculations is that if the relaxation rates $w$'s 
are comparable to the coupling constants $\Gamma$'s, then only a minor influence 
of higher excited states appears. They do not change the switching characteristics but increase the current amplitude.


%
As follows from the calculations in Sec. \ref{sec_phonon} the relaxation rates are of the order of a few GHz in the broad 
range of $V_{\rm QD}$. Therefore, one can easily obtain $\Gamma$'s smaller or comparable to the corresponding relaxation rates 
and consequently get the desired transistor behavior, as shown in the inset in figure \ref{currents_4levels}b.




\section{Current rectifier}\label{sec4}

Current rectification plays an important role in electron transport and is fundamental for the 
development of novel basic elements in nanoelectronics. The attempts have been made to build 
downscaled rectifiers. Coulomb blockade rectifier on a triple dot system \cite{stopa,yang} has been 
recently introduced. Rectification properties of a quantum wire coupled asymmetrically 
to a quantum dot have also been studied \cite{forchel}. We propose here a rectifier built on a DRN 
which utilizes different distribution of the ground and excited states wave functions.
	
To show the idea we discuss at first  a case where the  two lowest states are in the 
bias window  and we present the analytical solutions for the currents.

 \begin{figure}[h]
\begin{center}
\includegraphics[width=0.8\linewidth]{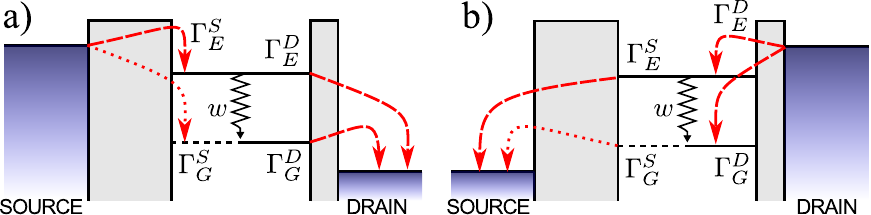}
\caption{The scheme of a rectifier with two energy levels in the bias window for forward (a)
and reverse (b) bias. The red dotted line represents the tunneling rate $\Gamma^S_G$ between the 
source electrode and the ground state that is assumed to be extremely small. The solid zig--zag 
line represent the relaxation $w$ from the excited state.}
\label{rectif}
\end{center}
\end{figure}

The confinement potential of the DRN is tuned so that the first excited state (E) has significantly 
larger overlap with the leads than the ground state (G). Thus, the same holds true for the corresponding 
tunneling rates, i.e, $\Gamma_G\ll \Gamma_E$. 
To get current rectification we have to break the left--right symmetry, therefore we assume  an 
asymmetric coupling to the source and drain electrodes, i.e., the drain electrode is much closer to the DRN 
than the source one. This results in the following relations between the tunneling rates:
\begin{equation}
\Gamma^S_i \ll \Gamma^D_i\ (i=G,E),
\label{gammas_rec}
\end{equation}
where $\Gamma^S_G$ is assumed to be so small that the current between the ground state and the source
electrode is negligible ($\Gamma^S_G \le 100$  kHz \cite{hans}). 
One can tune the distances between the DRN and electrodes to ensure the above relation. Moreover, 
it follows from equations in the Appendix that the ratio of the tunneling rates to the ground 
and excited states is independent of this distance, so that
\begin{equation}
\frac{\Gamma^S_G}{\Gamma^S_E}=\frac{\Gamma^D_G}{\Gamma^D_E},
\end{equation}
what is on line with the assumptions made, e.g., in Refs. \cite{Fudzi,Fudzi1} in the interpretation of 
their experiments.

The proposed mechanism of the rectification is the following: for forward bias 
($\mu_S-\mu_D>0$, see figure \ref{rectif}a) the 
current from the source 
electrode flows to the DRN through the excited state and then either further flows to the drain electrode 
through the excited state or the electron relaxes and leaves the DRN through the ground state. This is 
the forward direction of the rectifier. On the other hand, for reverse bias
($\mu_S-\mu_D<0$, see figure \ref{rectif}b) an electron can enter either the ground or the excited 
state. If the electron enters the ground state, it cannot leave the DRN
because of the negligible coupling  to the source electrode (we assume the temperature
to be low enough to prevent from exciting the electron to the next energy level and from tunneling back 
to the drain electrode). If the electron enters the excited state, it will relax to the ground state 
due to fast relaxation (Sec. \ref{sec2}).
Then, in either case the electron gets stuck in the
ground state blocking the current through the excited state by means of the Coulomb blockade. This is 
the reverse direction of the rectifier. 

To describe the proposed rectifier effect in a quantitative way we calculate the currents  
using the steady state solutions of the rate equations for the occupation probabilities [equations 
(\ref{matrix_eq}-\ref{current_i})].  Taking into account only the two lowest states we obtain 
 the  current $I_F$ for the forward bias 
and the reverse current $I_R$  for the reverse bias: 

\begin{eqnarray}
I_F&=&e\Gamma^S_E \: \frac{d_1\left(\Gamma^D_E + w\right)}{\Gamma^D_E + d_1w\left(1+\displaystyle\frac{\Gamma^S_E}{\Gamma^D_G}\right)}, \label{rectifier_eqs1}\\
I_R&=&-e\Gamma^S_G \: \frac{w + \Gamma^S_E}{w},
\label{rectifier_eqs2}
\end{eqnarray}
where $w$ is the relaxation rate for the first excited state and $d_1$ is its degeneracy.
These formulas allow us to  find the conditions under which DRN  behaves as a rectifier, i.e., for which
\begin{equation}
\left |\frac{I_{\rm F}}{I_{\rm R}} \right|\gg 1. 
\end{equation}
Since $\Gamma^S_E \ll w$  we get

\begin{equation}
\left|\frac{I_{\rm F}}{I_{\rm R}}\right|=\frac{\Gamma^S_E}{\Gamma^S_G} \: \frac{d_1\left(\Gamma^D_E + w\right)}{\Gamma^D_E + d_1w\left(1+\displaystyle\frac{\Gamma^S_E}{\Gamma^D_G}\right)}.
\label{relax}
\end{equation}

One can see that the ratio of the forward and reverse currents depends crucially on the ratio between 
couplings of the ground and excited states to the S electrode and this is a parameter that can easily be tuned
in a DRN.

We calculate the tunneling rates $\Gamma_E^S,\:\Gamma_G^S$ and $\Gamma_E^D,\:\Gamma_G^D$ and the relaxation
rates $w$ as a function of $V_{\rm QD}$ under the assumption that the geometry of the DRN is that the 
maximal value of $\Gamma_E^D$ is 5 GHz. 
\begin{figure}[h]
\begin{center}
\includegraphics[width=0.9\linewidth]{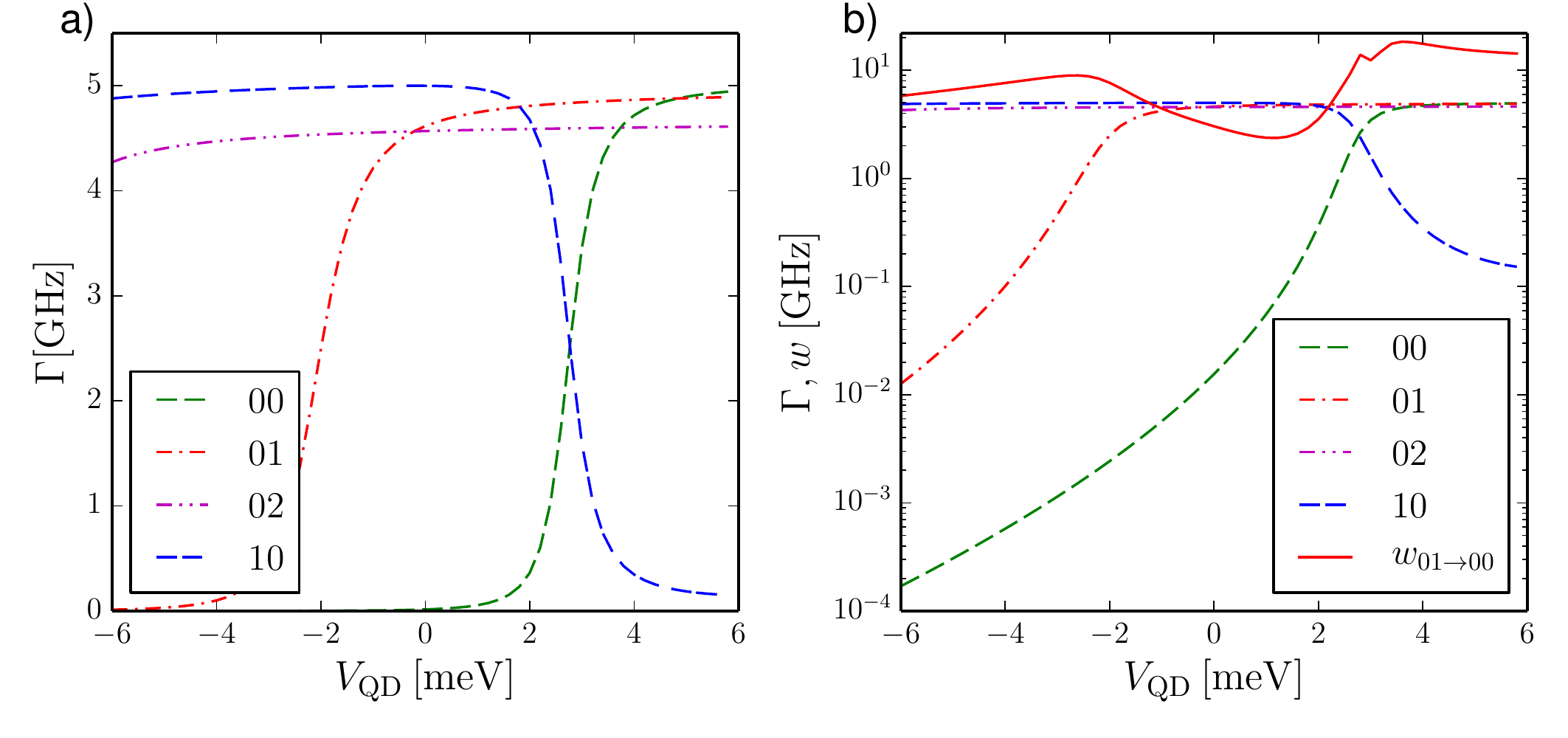}
\caption{The couplings $\Gamma$ between an electrode and particular states as functions 
of $V_{\rm QD}$ in linear (a) and logarithmic (b) scales. Additionally, the solid red line 
in the right panel represents the resulting relaxation rate.}
\label{gamma}
\end{center}
\end{figure}

Choosing $V_{\rm QD}=-4$ meV  (only two states in the bias window) for $\Gamma^D_E/\Gamma^S_E=100$ we get from 
equations (\ref{rectifier_eqs1}), (\ref{rectifier_eqs2}) and (\ref{relax})
  $I_F= 60$ fA, $I_R= 0.9$ fA, what gives $|I_{\rm F}/I_{\rm R}|=64$.

In this reference situation we get very small currents because both the G and E states are 
placed in the inner part of the DRN (see figure \ref{wavefunctions}a). To get a stronger current $I_F$ we have 
to choose larger $V_{\rm QD}$ 
for which some excited states are placed in the outer part of the DRN resulting in the larger ratio of the 
respective tunneling rates $\Gamma^S_{E_i}/\Gamma^S_G$. For $V_{\rm QD}>-2$ meV we have to consider three states 
in the bias window, as can be infered from figure \ref{fig_levels}. Below we present two examples of our numerical results:
\begin{table}[h]
\begin{center}
\begin{tabular}{cccccc}
\hline
$V_{\rm QD}$ & $\Gamma^D_E/\Gamma^S_E$ & $I_F$ & $I_R$ & $|I_F/I_R|$ & wave functions\\
\hline
\hline
1 meV & 100 & 9 pA & 0.09 pA & 100 & figure \ref{wavefunctions}b\\
0 meV & 200 & 4 pA & 0.012 pA & 320 & \\
\hline
\end{tabular}
\end{center}
\caption{Examples of model parameters and resulting currents.}
\end{table}


The wave functions in these two cases differ only qualitatively and therefore only the case of 
$V_{\rm QD}=1$ meV is presented in figure \ref{wavefunctions}b. 
The above example illustrates that it is possible to design a DRN that allows one to get a 
substantial forward current and a high degree of rectification.   
The Coulomb blockade due to the electron stuck in the ground state will prevent from transport 
through any of the excited states in the reverse direction. On the other hand,  they will all 
participate in the transport in the forward direction, what increases the total current. 
Because $\Gamma_S^G$ is not exactly equal to zero, we get some leakage 
current in the reverse direction but still substantial rectifying behavior occurs.
  
The crucial requirement for the rectifier is the strong difference between the coupling to the ground state 
and to the excited states. This is the point where the advantage of the DRN over a QD is clearly visible.
We performed, for comparison, calculations for QD with $R=70$~nm, $\Gamma^D_E/\Gamma^S_E=100$
and have got in the most favorable case:
 $I_F= 9$~pA, $I_R= 1.6$~pA, $|I_F/I_R|=5.7$.
One cannot obtain here a high degree of rectification because it is impossible to change the relative distribution
 of the ground and excited state wave functions in a QD. As a result, 
one cannot decrease the current $I_R$ without simultaneous decreasing of $I_F$.
 In QDs the ratio $\Gamma^S_E/\Gamma^S_G$ is usually below 10, 
whereas in the DRN the confining potential can be tuned to give $\Gamma^S_E/\Gamma^S_G$ up to $10^3$. We see
that the tunnel coupling to the reservoirs in the DRN is tunable over a much wider range than in QD. This, in turn,
results in larger values of currents and more efficient rectification.

\section{Summary}\label{sec6}

The fundamental requirement for future, low power consumption electronics is to control and manipulate single charges or 
spins. Quantum dots are the most popular and developed few--electron systems due to the relative easiness of fabrication 
and manipulation. On the other hand, the simple geometry of QDs allows modification of their electronic properties only 
to some extent, what encourages scientists to explore more complex systems like double or triple QDs. 
In this paper we performed systematic studies of electronic properties of a concentric dot-ring nanostructure and showed 
that, thanks to its non-trivial geometry, the structure offers unique possibilities to manipulate the electron wave 
functions. In particular, we have shown that by simple electrostatic gating one can move over the electron between the 
outer ring and the inner dot changing orbital relaxation by orders of magnitude and switching the character of a DRN 
from insulating to conducting. We have demonstrated that a DRN occupied by a single electron can be a good single 
electron transistor and a current rectifier with very high on/off ratio.
The presented so called {\it wave function engineering} technique allows designing the properties of a system from the 
lowest quantum mechanical level. This is exactly how modern nanotechnology works and is the way to reach the 
limits of device miniaturization and to make the next step in development of quantum computers.

In the paper we present results for a DRN build as a InGaAs structure.  However, the 
idea of a nanosystem where the wave functions can be moved over to different spatially separated parts can be applied also 
to other physical systems, for example graphene nanostructures 
\cite{graph-nano,graph-nano1,Trauzettel2007,Ponomarenko18042008}. We also neglect the spin effects. Yet, exploiting them
would allow one to use
a DRN in nanospintronics. The effect of spin on the Coulomb blockade has been studied 
in single QDs \cite{Johnson,Weis,Weinmann,hans} and in double QDs \cite{Ono,petta,simmons}. A spin blockade in single 
electron transistor in QD resulting from spin polarized leads has been discussed in \cite{HAWRYLAK}. In the case of a 
DRN one would be able to independly control spin--up and spin--down quantum states, what would lead to spin dependent 
tunnel rates $\Gamma$. This, in turn, would give the possibility to control a spin polarised current.

%
%
By studying one--electron properties we have demonstrated that 
the structure offers the unique possibilities to manipulate the distribution of the wave functions
what influences different DRN features. It also provides great flexibility in manipulating many--electron states,
what, however, is out of the scope of this paper and will be presented elsewhere.

\ack
This work was supported by the National Science Centre (NCN) grant DEC-2013/11/B/ST3/00824.
The authors thank K. Ensslin, T. Fujisawa, P. Hawrylak and J. Wr\'{o}bel for fruitful discussions. 

\appendix
\section{Derivation of the tunneling matrix element}

According to Bardeen's approach \cite{bardeen} the tunneling matrix element is given by
\begin{equation}
t_{\bm k}=\frac{\hbar}{2m^*}\int_S\vec{dS}\cdot\left[\psi_{\bm k}^*({\bm r})\vec{\nabla}\Psi_{nl}({\bm r})-\Psi_{nl}({\bm r})\vec{\nabla}\psi_{\bm k}^*({\bm r})\right],
 \label{app_1}
\end{equation}
what in two dimensions can be written as
\begin{equation}
t_{\bm k}=\frac{\hbar}{2m^*}\int^\infty_{-\infty}dy\left[\psi_{\bm k}^*({\bm r})\frac{\partial \Psi_{nl}({\bm r})}{\partial x}-\Psi_{nl}({\bm r})\frac{\partial \psi_{\bm k}^*({\bm r})}{\partial x}\right]_{x=x_0}, \label{app_2}
\end{equation}
where $\Psi_{nl}({\bm r})$ is the DRN's wave function and $\psi_{\bm k}({\bm r})$ is the  electrode's wave function.
The electrodes are modelled as half--planes and their wave functions and eigenenergies are given by
\begin{eqnarray}
\psi_{\bm k}({\bm r})&=&\frac{\sqrt{2}}{L}e^{ik_yy}\sin\left(\frac{\theta}{2}\right)e^{\kappa (x-x_1)}, \label{app_3}\\
\epsilon_{\bm k}&=&\frac{\hbar^2 k_y^2}{2m^*}+U-\frac{\hbar^2\kappa^2}{2m^*}, \label{app_4}
\end{eqnarray}
where
\begin{eqnarray}
\kappa&=&\frac{1}{\hbar}\sqrt{2m^*\left(U-\frac{\hbar^2k_x^2}{2m^*}\right)}, \label{app_5}\\
\sin\theta&=&\frac{2\kappa k_x}{\kappa^2+k_x^2}. \label{app_6}
\end{eqnarray}
The wave function of the DRN can be written as
\begin{equation}
\Psi_{nl}=R_{nl}(r)e^{il\phi}=R_{nl}(\sqrt{x^2+y^2})e^{il\arctan\frac{y}{x}}. \label{app_7}
\end{equation}
\newline
The derivatives in equation (\ref{app_2}) are given by
\begin{equation}
\frac{\partial \Psi_{nl}({\bm r})}{\partial x}=\left[\frac{dR_{nl}(r)}{dr}\frac{x}{\sqrt{x^2+y^2}}+R_{nl}(r)\frac{ily}{x^2+y^2}\right]\exp\left(il\arctan\frac{y}{x}\right) \label{app_8}
\end{equation}
and
\begin{equation}
\frac{\partial\psi_{\bm k}({\bm r})}{\partial x}=\kappa\frac{\sqrt{2}}{L}e^{ik_yy}\sin\left(\frac{\theta}{2}\right)e^{\kappa (x-x_1)}. \label{app_9}
\end{equation}
For each ${\bm k}$ in equation (\ref{app_1}) one has to calculate $\kappa$ and $\theta$ [equations (\ref{app_5}) and (\ref{app_6})], 
insert equations (\ref{app_3}), (\ref{app_7}), (\ref{app_8}), and (\ref{app_9}) into  (\ref{app_2}) and (numerically) calculate the integral.
It is convenient to express $t_{\bm k}$ in polar coordinates where it takes the following form
\begin{eqnarray}
t_{\bm k}&=&\frac{\sqrt{2}\hbar x_0}{2m^*L}\sin\left(\frac{\theta}{2}\right)e^{\kappa (x_0-x_1)}\int_{-\pi/2}^{\pi/2}d\phi\:
\frac{e^{i\left(k_yx_0\tan\phi+l\phi\right)}}{\cos^2\phi}\nonumber\\
&&\times
\left[R'_{nl}\left(\frac{x_0}{\cos\phi}\right)\cos\phi+R_{nl}\left(\frac{x_0}{\cos\phi}\right)
\left(\frac{il\sin2\phi}{2x_0}-\kappa\right)\right].
\end{eqnarray}

\section*{References}
\bibliography{duza}

\end{document}